\documentclass[showpacs,twocolumn,aps,jcp,superscriptaddress]{revtex4-1}
\usepackage[colorlinks, hyperfigures, linkcolor=black, citecolor=black, breaklinks,]{hyperref}
\usepackage[centertags]{amsmath}
\usepackage{amssymb,eqnarray}
\usepackage{amsthm}
\usepackage{fancyhdr}
\usepackage{graphicx,subfigure}
\usepackage{epsfig}
\usepackage{epsfig}
\usepackage{amssymb}
\usepackage{bbm}
\usepackage{graphicx}
\usepackage{amsmath}
\usepackage{subfigure}
\usepackage{bm}

\usepackage {CJK}

\newcommand*{\id}{{\normalfont\hbox{1\kern-0.15em \vrule width .8pt depth-.5pt}}}

\begin{document}
\title{Pseudo-chemotaxis of active Brownian particles competing for food}
\author{Holger Merlitz}
\email{merlitz@posteo.de}
\affiliation{Leibniz-Institut f\"{u}r 
Polymerforschung Dresden, Institut Theorie der Polymere, 01069 Dresden}
\author{Hidde D. Vuijk}
\affiliation{Leibniz-Institut f\"{u}r 
Polymerforschung Dresden, Institut Theorie der Polymere, 01069 Dresden}
\author{Ren\'e Wittmann}
\affiliation{Institut f{\"u}r Theoretische Physik II, Weiche Materie,
  Heinrich-Heine-Universit{\"a}t D{\"u}sseldorf, 40225 D{\"u}sseldorf, Germany}
\author{Abhinav Sharma}
\affiliation{Leibniz-Institut f\"{u}r 
  Polymerforschung Dresden, Institut Theorie der Polymere, 01069 Dresden}
  \affiliation{Technische Universit\"{a}t Dresden, Institute of Theoretical Physics, 01069 Dresden, Germany}
\author{Jens-Uwe Sommer}
\affiliation{Leibniz-Institut f\"{u}r 
Polymerforschung Dresden, Institut Theorie der Polymere, 01069 Dresden}
\affiliation{Technische Universit\"{a}t Dresden, Institute of Theoretical Physics, 01069 Dresden, Germany}


\begin{abstract}
  Using Brownian dynamics simulations, the motion of active Brownian
  particles (ABPs) in the presence of fuel (or 'food') sources is studied.
  It is an established fact that within confined stationary systems,
  the activity of ABPs generates density profiles that are
  enhanced in regions of low activity, 
  which is generally referred to as 'anti-chemotaxis'.
  We demonstrate that -- contrary to common believes -- in non-stationary
  setups, emerging here as a result of short fuel bursts, our model
  ABPs do instead exhibit signatures of chemotactic behavior.
  In direct competition with inactive, but otherwise identical Brownian
  particles (BPs), the ABPs are shown to fetch a larger amount of food.
  From a biological perspective, the ability to turn active would,
  despite of the absence of sensoric devices, encompass an
  evolutionary advantage. 
\end{abstract}
\maketitle

\section{Introduction}

The ability to move is among the common features of living
systems. Mobility opens the opportunity to escape from dangerous
environments and to search for food
sources to increase the intake of energy-rich substances, which
enhance the metabolism, grant survival and eventually the chance
for reproduction -- the central theme of the game of evolution. 
It is not surprising that mobile lifeforms are abundant on all
scales including the simplest microscopic organisms.
Prominent examples are white blood
cells chasing intruders~\cite{fenteany_COH04}, sperm cells
rushing toward the oocyte~\cite{Pfeffer_84} or bacteria
such as Escherichia coli searching for food~\cite{berg_04}.

Such a motion involves a delicate apparatus of chemical sensors,
information processing and direction control and is commonly
called chemotaxis~\cite{berg_93}. The question then arises how
primitive organisms came to 'learn' the mechanisms required for
chemotaxis during early stages of evolution.  In this context,
recent theoretical and simulation studies of transport phenomena
in systems with active Brownian particles
(ABPs)~\cite{bechinger_RevModPhys16,yamamoto_powder15}, i.e.\ lifeless
self-propelling micro- or nanomotors, can provide valuable insights: 
It has been demonstrated that these ABPs are capable of running
up a spatial activity gradient under certain, non-stationary
conditions~\cite{peng_AC15,ghosh_PRE15,vuijk_PRE18}. Since this
phenomenon does not involve any mechanism of sensing or data
processing, the term {\it pseudo-chemotaxis} has been coined
to contrast the same from ordinary chemotaxis, which, in
addition to space-dependent activity, also requires a coupling
between the direction of the fuel gradient and the orientation
of the particle~\cite{popescu_NanoLett18,lozano_NatureComm16}.
Even though such a coupling has been shown to exist
for torque-free spherical ABPs and to
affect their motion in activity
gradients~\cite{sharma_PRE17,merlitz_JCP18}, 
the involved orientation bias is generally very low
and actually directed against the activity
gradient~\cite{vuijk_PRE18}, i.e.\ reducing the impact
of pseudo-chemotaxis. The fundamental mechanism behind
pseudo-chemotaxis has been described in terms of a dynamical
reflection of the particle trajectory on the boundary between
regions of different activity levels and shown to be vastly
enhanced if the particle is allowed to store a single bit
of memory~\cite{kromer_preprint19}.

In the laboratory, synthetic self-propelling agents, notably
Janus-particles of spherical shape, are designed and
employed to study the features of simple self-propelling
motors. These particles are driven by catalytic reactions
with a supplement to the solvent such as hydrogen
peroxide~\cite{howse_PRL07} or hydrazine~\cite{gao_JACS14},
or driven by thermal forces that are induced with intense
light~\cite{lozano_NatureComm16}. The smallest Janus-particles
are of dimensions of only $30$nm and thus of the same order
as some of the proteins found in living systems~\cite{lee_NanoLett14}.
As a matter of fact, catalytic enzymes have been reported
to exhibit enhanced diffusion during catalytic
activity~\cite{muddana_JAmChemSoc10,sengupta_JAmChemSoc13,riedel_Nature15},
and the enhanced mobility has recently been claimed to originate from a
(so far unknown) mechanism of self-propulsion~\cite{jee_PNAS18}.
This would imply that there exist biological systems, far
simpler than living cells or bacteria, which feature
self-propelled motion on a similarly low level of complexity
as synthetic Janus-particles.


In the present work, we pick up previous studies of ABPs
in activity gradients~\cite{vuijk_PRE18} and -- motivated by
biological evolution -- attempt to approach
the phenomenon of pseudo-chemotaxis from another angle:
While transport properties
such as first passage times and target hit probabilities have
previously been studied in the presence of external activity fields,
the present work interprets activity as a result of the intake of
explicit food (or fuel) particles of limited resources in a
competitive environment. Two species, ABPs, which
temporarily turn active after food intake, and identical Brownian
particles (BPs), which consume food but remain inactive, are
competing for the food available in the system. In Sec.\
\ref{sec:model} we describe the numerical model that was
implemented for the following Brownian dynamics simulations.
Section \ref{sec:confined} describes stationary confined systems
of ABPs and BPs into which food is injected continuously.
We show that in such a setup the activity leads to an
anti-chemotaxis as commonly reported for ABPs. In Sec.\
\ref{sec:open} the systems are open and food is injected
in terms of a short burst that emerges at a given distance
to the particles.
Here, pseudo-chemotaxis is observed and the ABPs exhibit
an increased food consumption when compared to the passive
BPs. Section \ref{sec:gradients} investigates different food
distributions and demonstrates that gradients in the food
concentration are necessary for the ABPs to gain a significant
advantage over their passive competitors. Our findings are
summarized and discussed in Sec.\ \ref{sec:conclusion}.

\begin{figure*}[t]
\includegraphics[width=1.5\columnwidth]{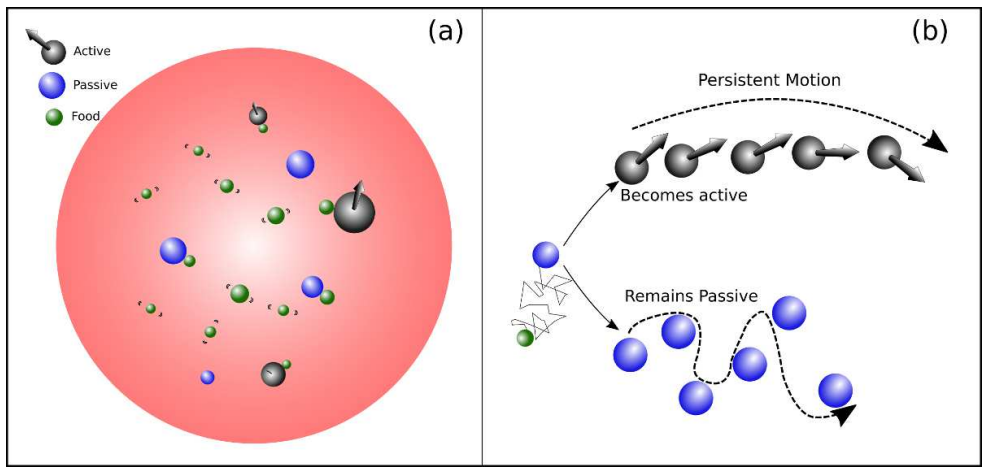}
\caption{Schematic of particle characteristics, left panel: The system
  contains passive BPs (blue), ABPs (grey) and food particles
  (green). If an ABP runs into a food particle, it temporarily turns
  active and exhibits a persistent motion (right panel). The BP
  consumes food, too, but remains a passive Brownian walker. Which
  strategy is superior?
}
\label{fig:schematic}
\end{figure*}

\section{Numerical Model}\label{sec:model}

Two types of particles exist (see Fig.\ \ref{fig:schematic}):
First, the standard Brownian
particles (BPs) and, second, self-propelling active Brownian
particles (ABPs), activated for a given time-period,
called boost time $\tau_{\rm B}$, after consuming a
dimensionless (point-like) food particle.
The spherical particles have diameters of $d = 30$nm and
are dispersed in an implicit solvent of temperature
$T = 298$K and dynamic viscosity  $\eta = 0.89$mPa$\cdot $s
(corresponding to water at 298K). The resulting passive
(i.e.\ in absence of self-propulsion) translational
diffusion coefficient amounts to $D_{\rm t} = 0.016$nm$^2$/ns,
and the rotational relaxation time is of the order of
$\tau_{\rm r} = 9.2 \cdot 10^3$ns.


ABPs -- when active -- are driven with a force of $f = 1$pN,
directed along the axis of their current orientation. This
driving force generates an additional contribution to the translational
diffusion of $D_{\rm a} = 0.048$nm$^2$/ns, thrice the value
of their passive diffusion. The driving force of $1$pN is
of the same order as recently proposed forces acting on
catalytic enzymes of similar sizes~\cite{jee_PNAS18}. We
assume that the activity does not affect the rotational
diffusion of the ABP. The point-like food particles are
assumed to move with a translational diffusion coefficient
that is larger by a factor five when compared to the BPs.
A food particle is 'consumed', i.e.\ removed from the system,
as soon as it occupies the space taken by another particle.

Both BPs and ABPs may consume an unlimited number of food particles
at any time (being active or not). After consumption of each
food particle, the ABPs remain active only for a fixed
time period, called boost time $\tau_{\rm B}$, 
i.e.\ no accumulation and storage of energy beyond a single
quantity is taken into account. 

Since we are interested in the dynamics of single particles,
pair-interactions between particles as well as hydrodynamic
interactions are absent. In confined systems, all particles
(including food) interact with the confining walls via a
standard repulsive Weeks-Chandler-Andersen-potential~\cite{weeks_JCP71}.
The time integration is
carried out using a standard second-order Brownian dynamics
algorithm~\cite{klenin_BiophysJ98} with a time step of
$dt = 10$ ns. The system parameters are summarized 
in Table \ref{tab:quantities}.

\begin{table}[htb]
\renewcommand{\arraystretch}{1.1}%
\begin{tabular}[b]{@{}r@{}l|c@{}}
\multicolumn{2}{@{}c|}{Quantity} & Meaning \\
\hline 
$d$         & $\,=3$       & dimension of space    \\  
$b$       & $\,= 30$ $nm$      & particle diameter      \\
$m$       & $\,= 14.14$ $ag$       & particle mass             \\
$dt$        & $\,= 10$ $ns$   & simulation timestep    \\
$T$         & $\,= 298$ $K$     & system temperature     \\
$f$         & $\, = 1$ $ag\cdot nm / ns^2$ & driving force (ABP)\\
$\eta $      & $\, = 0.981$ $mPa \cdot s$ & viscosity
(solvent)\\
$\zeta = 3 \pi \eta b$ &$\,= 251.9$ $ag/ns$ & frictional drag coeff. \\
$\tau_m = m/\zeta$   & $\,=0.0561$ $ns$  & momentum relax.\ time \\ 
$D_t = k_BT/\zeta$&$\,=0.0163$ $nm^2/ns$ & passive diffusion coeff.\  \\
$D_r = 3 D_t / b^2$  & $\,=5.45\cdot 10^{-5}$ $ns^{-1}$ & rot.\ diffusion
coeff.\ \\
$\tau_r = 1/(2D_r)$ &$\,=9180$ $ns$ & rot.\ relaxation time       \\     
$v = f/\zeta$ & $\, = 3.97\cdot 10^{-3}$ $nm/ns$ & final velocity (ABP)  \\
$D_a= v^2\tau_r/d$ & $\, = 0.0482$ $nm^2/ns$ & activated diffusion coeff.\ \\
$T_a= \zeta D_a /k_B$& $\,= 880$ $K$ & activated temperature\\
\end{tabular}
\caption{Simulation parameters and derived quantities.
\label{tab:quantities}}
\end{table}

\section{Results}
\subsection{Confined and stationary systems}\label{sec:confined}

\begin{figure}[t]
\includegraphics[angle=270,width=\columnwidth]{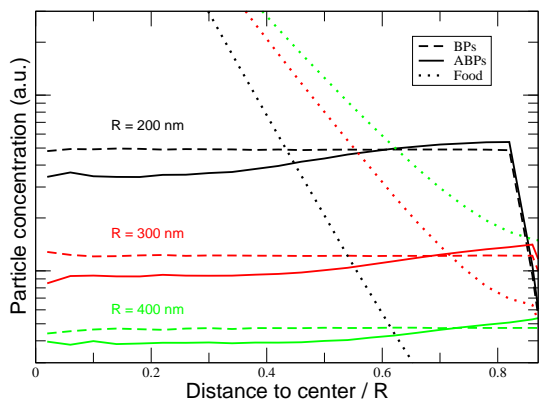}
\caption{Particle concentration as a function of (relative)
  distance to the center, for different sizes $R$ of the
  reactor. Close to the food source, ABPs are active most
  of the time, leading to a significant depletion of their 
  concentration. Close to the outer wall (right edge of
  the graphic), particle concentrations
  drop due to steric repulsion. Food-production rate: $50/\mu$s; number
  of particles: 100 (of each species, ABPs and BPs);
  simulation time: $2$s; boost time of ABPs: $\tau_{\rm B} = 2 \tau_{r}
  \approx 18 \mu$s.}
\label{fig:radial_stationary}
\end{figure}

In a first set of simulations, a total of 200 particles
(100 of each species, BPs and ABPs) were placed into 
spherical confinements of radii $R \in \{200, 300, 400\}$
nm. At the center, food particles were randomly
generated at an average rate of $50$ particles/$\mu$s.
After a sufficiently long simulation time, all particle
distributions turned stationary and are displayed in
Fig.\ \ref{fig:radial_stationary}.

The stationary distributions of fuel-activated ABPs are non-uniform, being
reduced in regions in which the food concentration is high (i.e.\
close to the center of the confinement). This is a
consequence of the self-propulsion that comes along with their
food consumption: It is well known that the stationary concentration of an ABP
is inversely proportional to its average driven velocity
and thus highest in regions of diminishing activity~\cite{schnitzer_PRE93,merlitz_SM17}.
In our simulations, this is visible at all system sizes, though the quantity
of the ABP-depletion is highest in the small system
($R = 200$nm, solid black curve, note the logarithmic scale),
while the distributions of BPs naturally remain unaffected
by the food concentrations. The phenomenon observable
in these simulations is thus the well known anti-chemotaxis
of ABPs in stationary systems with activity gradients.

\begin{figure}[t]
\includegraphics[angle=270,width=\columnwidth]{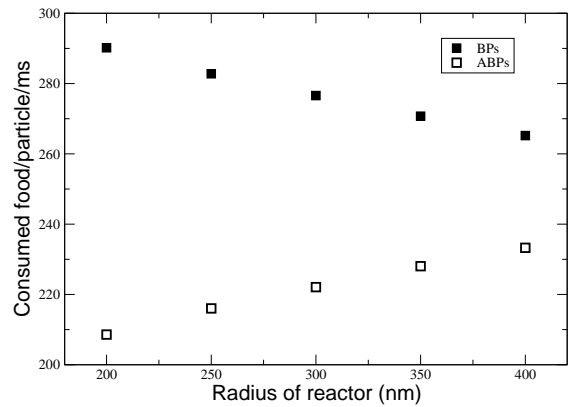}
\caption{Food consumption rates of particles for different sizes
  of the reactor. As a result of the bias of their density distributions,
  ABPs consume less food that BPs, exhibiting anti-chemotaxis.
  With increasing size of the reactor, the ratio of activated
  to inactive APBs is diminishing and their food consumption rates
  are approaching those of the BPs.}
\label{fig:food_stationary}
\end{figure}

Figure \ref{fig:food_stationary} displays the average food
consumption of each species in systems of different radii:
In each case, the passive BPs are able to fetch food at
a higher rate when compared to the ABPs, as a result of
the reduced presence of ABPs close to the food source.
 We have repeated the simulations
with different parameter settings, varying boost
times $\tau_{\rm B}$ and food production rates, 
arriving at the same qualitative conclusions:
Anti-chemotaxis reflects itself in the average
food consumption of the active particle species, and
any degree of activity may be regarded a disadvantage
as soon as food is needed for the metabolism of a
fictitious proto-lifeform. However, the validity of
this conclusion is restricted to stationary systems,
achieved by a continuous and infinite supply of food,
as is going to be demonstrated in the following section.

\subsection{Open and non-stationary systems}\label{sec:open}

In the following set of simulations, we mimic a situation
which may occur frequently in natural environments such as
open bodies of water, e.g.\ ponds or lakes: A temporary
food source emerges and several species
in its vicinity start to compete for food. The system is
unconfined (on relevant length-scales) and the distributions
of different species are non-stationary.
Once again, we compare the average food consumption of
passive BPs and ABPs.

Sets of 100 representatives of each species are initially
randomly distributed on a shell of distance $r_{\rm ini}$ to
a food source, which emits a given number of (here: 5000)
food particles in an instant. These food particles
subsequently diffuse into all directions, run into the
surrounding particles and are eventually consumed. The ABPs are activated
for a boost time of $\tau_{\rm B} = 2 \tau_{\rm r} \approx 18 \mu$s
upon consumption of a food particle, and this activity affects
their transient concentration profiles.

\begin{figure}[t]
\includegraphics[width=\columnwidth]{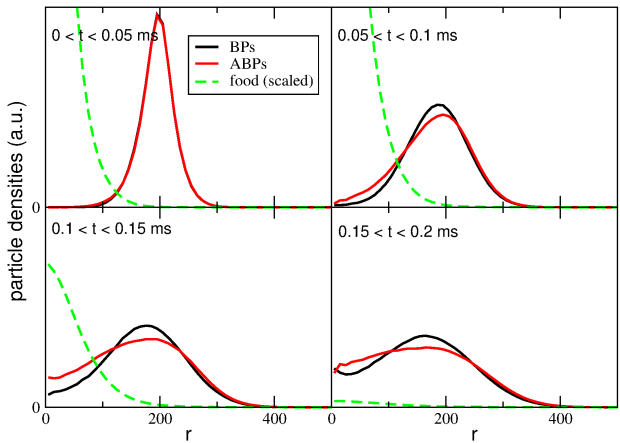}
\caption{Time evolution of the density distributions of BPs (black)
  and ABPs (red), as well as food-particles (green, scaled down),
  for different time intervals.
  Initially, BPs and ABPs start at the distance $r_{\rm ini} = 200$nm to
  the center, boundaries are absent. A burst of food particles
  occurs at the center at $t = 0$. Upon food consumption,
  the density distribution of ABPs advances rapidly toward the food
  source. Boost time of an ABP after consumption of a food-particle:
  $\tau_{\rm B} = 2 \tau_{\rm r} \approx 18 \mu s$.
}
\label{fig:time_evolution}
\end{figure}

\begin{figure}[t]
\includegraphics[angle=270,width=\columnwidth]{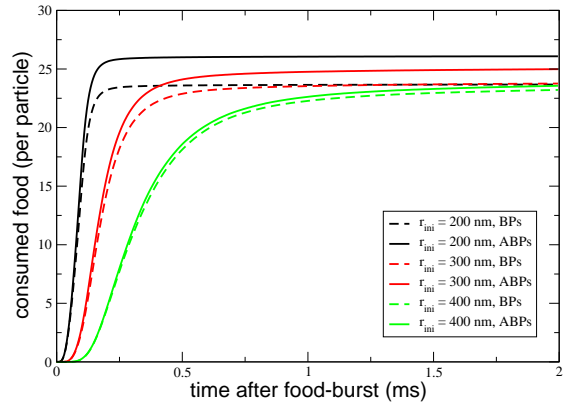}
\caption{Consumed food particles as a function of time after begin of
  the food burst. As a result of the bias of their density distributions,
  ABPs consume more food that BPs, exhibiting chemotaxis. The differences
  diminish with increasing initial distance $r_{\rm ini}$ to the center, since
  fewer ABPs are activated.
}
\label{fig:food_consumption_spills_t2}
\end{figure}

The time evolution of these profiles is displayed in
Fig.\ \ref{fig:time_evolution} for the case of
$r_{\rm ini} = 200$nm, for different time intervals
(chosen to keep the statistical noise low). The initial
$\delta$-distributions
of both particle species begin to spread and to run
into the expanding distribution of food particles (panel
a). Since the ABPs begin to propagate at an enhanced speed,
the tail of the corresponding concentration profile (red) reaches
out toward the center of the food source at which food
concentration is highest (panels b-d). When compared to the
passive BPs (black profiles), the ABPs reach 
regions of high food concentration at an earlier time.
This advantage is reflected in
the rates of food consumption as shown in Fig.\
\ref{fig:food_consumption_spills_t2} for three different
initial distances of the particles: In particular at short
distance to the food (black curves), the ABPs (solid curve)
are able to fetch a larger number of food particles than
the BPs (dashed curve).

The differences diminish with
increasing initial distances of the particles ($r_{\rm ini} = 300$nm,
red curves, $r_{\rm ini} = 400$nm, green curves), since here
the food dilutes to a degree that the ABPs remain increasingly
deactivated. Once the boost time $\tau_{\rm B}$ is increased so
that an ABP profits from elongated periods of activation, its
advantage over the BP increases even in the case of larger
initial distances to the food source (data not shown).

In a non-stationary setup, ABPs are therefore exhibiting a kind
of chemotaxis which enables them to consume a larger amount of
food than their non-active competitors. We call this
phenomenon {\it pseudo-chemotaxis} because it does not involve
any sensing of food gradients on the part of the active particle.
In fact, no orientation-bias is required for 
pseudo-chemotaxis, as was already shown in a previous
work~\cite{vuijk_PRE18}: While ABPs do even exhibit a small degree of
orientation-bias inside activity gradients, this bias is pointing
down the gradient and thus working against chemotaxis,
yet remaining sufficiently weak to be negligible in the systems
considered here.

\subsection{The impact of food/fuel gradients}\label{sec:gradients}

In order to analyze the importance of food concentration
gradients, two test setups with immobile food particles
are compared: First, the food has a Gaussian concentration:
being allowed to diffuse from the location of their production
for a short time and subsequently immobilized. In the second
setup, food is distributed uniformly within a spherical
volume of radius $r = 100$nm. Both setups contain 5000 food
particles, while the remaining particle species start at
$r_{\rm ini} = 200$nm, outside the food sources, and the
systems are unbounded.

\begin{figure}[t]
\includegraphics[width=\columnwidth]{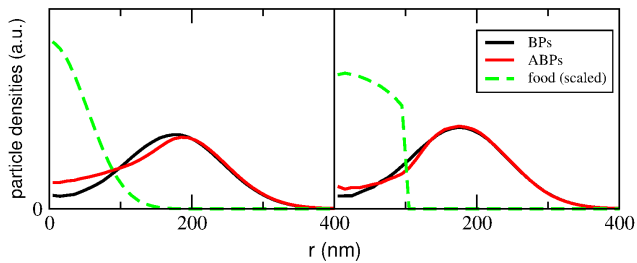}
\caption{Time evolution of the density distributions of BPs (black)
  and ABPs (red), as well as food-particles (green, scaled down),
  averaged over the time interval of $0.1 < t < 0.15$ms.
  Initial distance of the particles to the food source:
  $r_{\rm ini} = 200$nm. Left panel: initial food profile
  is Gaussian, right panel: Initial food profile is
  a step-function. Food particles are immobile.}
\label{fig:time_evolution_2}
\end{figure}

While the particle distributions spread, a certain fraction enters the
food-reservoirs and begins to consume food. ABPs are then
turning active and once again this activity affects the
dynamics of their concentration profiles. Figure \ref{fig:time_evolution_2}
displays typical snapshots of the concentration profiles,
averaged over the time interval of $0.1 < t < 0.15$ms after
setoff. The system that features a food gradient (left panel)
once again exhibits an increased concentration of ABPs close
to the food source at which the food density is at its maximum.
Contrary to that, the (initially) uniform food distribution
leads to a different situation (right panel): Although close
to the center of the food source, there exists an enrichment of
ABPs compared to BPs, the situation is the opposite closer to
the periphery of the food reservoir, at which BPs are
more abundant than ABPs. Since the food distribution is at
this moment still fairly uniform, the enrichment at the center
does not encompass any significant advantage in terms of food
intake.

\begin{figure}[t]
\includegraphics[angle=270,width=\columnwidth]{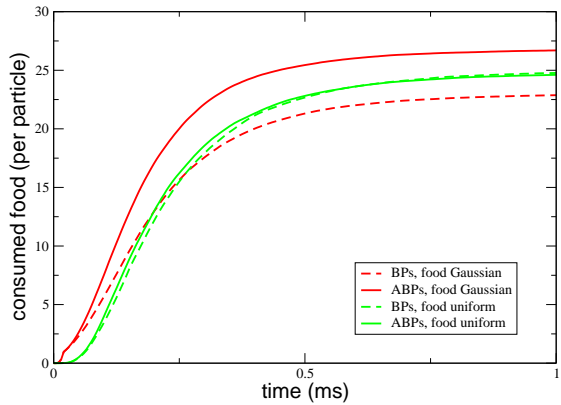}
\caption{Consumed food under the condition that the food
  concentration profile is a step function (green) or a
  Gaussian function (red). In the latter case, the ABPs
  experience an advantage over the passive BPs in terms of
  the amount of consumable food. 
}
\label{fig:food_consumption_different_setups}
\end{figure}

Figure \ref{fig:food_consumption_different_setups} exhibits
that both setups in fact lead to entirely different outcomes
in terms of the food consumption: In the presence of a food
gradient, the ABPs are able to fetch a higher number of food
particles than passive BPs (red curves). Starting with a uniform
food distribution does not lead to any systematic advantage for the ABPs
(green curves). We have varied system parameters such a $r_{\rm ini}$
or the boost time $\tau_{\rm B}$ of ABPs without arriving at
qualitatively different outcomes: Whenever a gradient existed
in the food concentration, the ABPs were able to take profit
and gained a larger amount of food, while in the absence of such
a gradient both BPs and ABPs received close to identical results.
The phenomenon of pseudo-chemotaxis of ABPs in open, non-stationary
systems is thus greatly supported by gradients which lead to high
food/fuel concentrations.

\section{Conclusion}\label{sec:conclusion}

In the present work, we have approached the phenomenon of
pseudo-chemotaxis of ABPs -- recently discussed in the
framework of transport properties such as first passage
times and target hit probabilities~\cite{ghosh_PRE15,vuijk_PRE18} --
from another angle that is motivated by evolutionary
considerations. The question that had to be addressed was: Does
pseudo-chemotaxis actually lead to an advantage for a particle
that is capable of food-induced motion, but unable to sense
food gradients?

In our simulation study, two particle species had to compete
for limited resources of food, the latter being simulated
in terms of explicit food particles that could be consumed
to trigger a temporary state of self-propulsion (in case of
ABPs). The simulation parameters were chosen to be
of similar dimensions as observed in real systems such as
Janus particles of $30$nm diameter~\cite{lee_NanoLett14} or 
catalytic enzymes, which appear to turn active during enzymatic
activity~\cite{jee_PNAS18}.   

The simulations confirm the well known fact
that within a confined system and after reaching
stationary, the ABPs preferably occupy
regions of low activity. Consequently, our fuel-activated
ABPs do also consume a smaller
amount of food than (otherwise identical) passive BPs
(Sec.\ \ref{sec:confined}). This is an unmistakable
signature of anti-chemotaxis that has frequently been reported 
for ABPs in systems with activity gradients.

Non-stationary setups however can lead to the opposite
situation in which ABPs exhibit an advantage over BPs
(Sec.\ \ref{sec:open}). The term pseudo-chemotaxis
has been coined because -- contrary to ordinary chemotaxis --
it does not involve any features related to sensing or
information processing. The ABPs are running up activity
gradients  as a result of a purely statistical effect,
related to their enhanced mobility in these regions,
and the present work demonstrates that the observed bias in
their transient concentration profile does in fact
lead to an advantage over non-active competitors in terms
of food of fuel intake. In Sec.\ \ref{sec:gradients} we further
demonstrated that not only the ability to turn active,
but also the presence of a food concentration gradient
was necessary for the ABPs to surpass the
food consumption of their non-active competitors.

From a biological point of view it is certainly
tempting to speculate about a potential relevance of these insights
for the evolution of early proto-lifeforms, which were sufficiently
complex to feature a mechanism for self-propulsion, but yet unable
to perform any kind of sensing or data processing. It would be
astonishing if the feature of pseudo-chemotaxis would not have
made it into the toolkits of early evolution. Catalytic enzymes
may be modern examples for complex molecules which exhibit such
a combination of features. Pseudo-chemotaxis might thus
have been a vehicle -- based entirely on non-equilibrium physical
processes -- which enabled similar entities to enter the game
of evolution, to compete for fuel or food, the potential to enhance
their metabolism, and eventually their reproduction rate.

As a next level of realism, food consumption could be coupled
to the reproduction rates of species, while the energy required for
self-propulsion could be accounted for. Pair-potentials between
different species and food particles could be implemented to
generate correlated dynamics and mimic predator-prey interactions,
which might also include a minimalist model for information
processing.



\end{document}